\documentclass{article}
\usepackage[utf8]{inputenc}
\usepackage{braket}
\usepackage{amsmath}
\usepackage{graphicx}
\usepackage{enumitem}   
\usepackage{mathrsfs}
\usepackage{dsfont}
\usepackage{mathtools}
\usepackage{amssymb}
\usepackage{xcolor}
\usepackage{physics}
\usepackage{comment}
\usepackage{hyperref}
\usepackage{authblk}

\DeclareMathAlphabet{\pazocal}{OMS}{zplm}{m}{n}

\usepackage[square,numbers]{natbib}

\newcommand{\zerodel}{.\kern-\nulldelimiterspace}

\renewcommand\bra[1]{{\left<#1\right|}}
\renewcommand\ket[1]{{\left|#1\right>}}
\newcommand{\kket}[1]{\left|\left\zerodel  #1 \right> \right>}
\newcommand{\bbra}[1]{\left< \left<  #1  \right|\right\zerodel}
\newcommand{\brakket}[2]{\left\zerodel\left<  #1  \left|   #2  \right> \right>\right\zerodel}
\newcommand{\bbrakket}[2]{\left\zerodel\left< \left<   #1 \left|   #2  \right> \right>\right\zerodel\right\zerodel}
\newcommand{\bbraket}[2]{\left< \left<   #1 \left|  #2  \right>\right\zerodel\right\zerodel}
\newcommand{\eexpval}[1]{\left< \left<  #1 \right> \right>}

\newcommand*\samethanks[1][\value{footnote}]{\footnotemark[#1]}

\title{Measuring time in a timeless universe}
\author[1]{Sam Kuypers\thanks{These authors contributed equally to this work.}\thanks{samuelkuypers@gmail.com}}
\author[2]{Simone Rijavec\samethanks[1]\thanks{simone.rijavec@physics.ox.ac.uk}}
\affil[1]{\normalsize{Département d'informatique et de recherche opérationnelle,} \linebreak
\normalsize{Université de Montréal, Canada}}
\affil[2]{Clarendon Laboratory, University of Oxford, Parks Road, Oxford OX1 3PU, United Kingdom}

\date{\today}

\begin{document}
\maketitle

\begin{abstract}
Physical systems are usually assumed to evolve relative to an external time parameter, which is problematic because in quantum theory that parameter is not a physical observable. Page \& Wootters (1984) solved this by proposing that the universe is in a stationary state, eliminating the need for the external time parameter.  Instead, their model contains an isolated subsystem, a ‘clock’, with which other subsystems are entangled, making the latter appear to evolve relative to different states of the clock.  While this resolves the problem of the time parameter, the assumption that the clock is isolated prevents it from being measured, as this requires an interaction with another system. We prove that the clock can be measured while preserving the core features of the Page-Wootters construction. We also discuss clock synchronisation.
\end{abstract}

\section{Introduction}
Time has a peculiar status in the foundations of physics: it is typically invoked in dynamical theories as a real-valued parameter, \(\lambda\), relative to which systems can change, yet this time parameter is not physical. For instance, in quantum theory, \(\lambda\) is not an observable of any system. It cannot be one because the observables of quantum systems are matrix- or operator-valued (in the terminology of Dirac \cite{dirac1981principles}, such observables are \textit{q-numbers}), whereas \(\lambda\) is a real-valued parameter. This difference is especially striking in experimental physics since experimental physicists will use quantum systems as timekeeping devices or \textit{clocks} during experiments, an example being the atoms in an atomic clock. These quantum systems possess q-number observables, highlighting a fundamental discrepancy, namely that the time parameter \(\lambda\) cannot be equivalent to any of the observables of the quantum systems that are used as clocks.

This mismatch between fundamental theories of physics and practical reality is problematic; it is one of the key conceptual issues in our understanding of time – for further discussions of related conceptual difficulties, usually referred to as `the problem of time' in quantum gravity,  see Refs.~\cite{barbour2009nature, fiscaletti2015timeless,kuchar_time_2011,isham_canonical_1993}.\footnote{Another aspect of the problem of time is that the time parameter provides a circular explanation of change. In conventional physics, physical systems change because time changes, but as an account of change this is unsatisfactory because it leaves unexplained why the time parameter itself changes.} This problem is resolved in `timeless' approaches to physics, which have as their overarching theme that time is not fundamental but emergent. In these constructions, time emerges from more fundamental attributes of static physical systems; one advantage of this view is that it ensures time is always a physical quantity, and such `timeless' theories often provide straightforward ways of measuring time.

A notable example of a timeless approach in classical physics is a model due to Barbour \cite{barbour2009nature}. Barbour's model consists of point masses interacting via Newtonian gravity in such a way that their configurations in space are unique at any instant (their uniqueness is guaranteed by the conservation laws Barbour imposes), implying that a particular instant of time can also be defined by its unique configuration of masses. Thus, in Barbour's model, time is not an external parameter, like \(\lambda\), but an internal, measurable attribute, making \(\lambda\) superfluous so that it can be dispensed with.\footnote{Barbour expands on these ideas in \textit{The end of time} \cite{barbour2001end} and in his more recent work on a theory called shape dynamics \cite{barbour2014solution}.}

\section{The Page–Wootters construction} \label{PWconstruction}

In this paper, we shall explore a timeless approach in quantum theory known as the  Page–Wootters construction \cite{page1983evolution}, which explains time as an entanglement phenomenon. Here, we provide a brief description of the Page–Wootters construction, based on modern understandings of it, such as those presented in Refs.~\cite{giovannetti2015quantum,marletto2017evolution}. 

The Page–Wootters construction is motivated by the superselection rule for energy, which prohibits coherent superpositions of the eigenstates of the Hamiltonian, \(\hat{\mathcal{H}}\). As such, we make the following two assumptions: first, for simplicity, we assume that the universe is in a pure state, described at the initial time \(\lambda=0\) by a state vector \(\kket{\Psi}\), which we will henceforth refer to as the \textit{universal state vector}; second, to ensure compatibility with the superselection rule, we assume that the universal state vector is an eigenstate of \(\hat{\mathcal{H}}\) with eigenvalue zero:
\begin{equation} \label{eq:stationary}
\hat{\mathcal{H}}\kket{\Psi} = 0,
\end{equation}
where \(\kket{\cdot}\) indicates that a vector resides in the full Hilbert space rather than merely in one of its subspaces. Note that the zero-energy assumption is merely convenient, as the construction works for any eigenvalue of \(\hat{\mathcal{H}}\) \cite{giovannetti2015quantum}.

The universe described by \(\kket{\Psi}\) is a closed system, so it evolves unitarily according to the unitary operator \(e^{-i\hat{\mathcal{H}}\lambda}\), where \(\lambda\) is the time parameter. This means that if \(\kket{\Psi}\) is the state of the universe at time \(\lambda = 0\), then its state at a general time \(\lambda \) is given by \(e^{-i\hat{\mathcal{H}}\lambda}\kket{\Psi}\). An energy eigenstate is a stationary state so that
\(e^{-i\hat{\mathcal{H}}\lambda} \kket{\Psi} = \kket{\Psi}\), as follows from from Eq.~\eqref{eq:stationary}. Therefore, the assumption that the universe is in an eigenstate of \(\hat{\mathcal{H}}\) ensures that no measurable quantity depends on \(\lambda\). Consider, for example, the expectation value of an arbitrary observable \(\hat{A}\) at an arbitrary time \(\lambda\), which is equal to its expectation value at \(\lambda = 0\) since
\begin{equation}
\bbra{\Psi}e^{i\hat{\mathcal{H}}\lambda}\hat{A}e^{-i\hat{\mathcal{H}}\lambda}\kket{\Psi} = \bbra{\Psi}\hat{A} \kket{\Psi}.
\end{equation}
Evidently, the expectation value of any observable at any time \(\lambda\) is equivalent to its expectation value at \(\lambda =0\), so the time parameter might as well not be there at all.

How, then, can it be that there appears to be evolution and change in the real universe? To explain this, Page \& Wootters assumed that the universe can be divided into two non-interacting subsystems: a clock, \(\mathcal{C}\), with Hilbert space $\mathscr{H}_\mathcal{C}$, and the rest of the universe, \(\mathcal{R}\), with Hilbert space $\mathscr{H}_\mathcal{R}$, where the latter system might in turn consist of further subsystems.  \(\mathcal{C}\) and  \(\mathcal{R}\) jointly constitute the entire universe so that the Hilbert space of the universe is $\mathscr{H}_U = \mathscr{H}_\mathcal{C} \otimes \mathscr{H}_\mathcal{R}$. Because we assume that \(\mathcal{C}\) and \(\mathcal{R}\) are non-interacting, the Hamiltonian of this model must be a sum of the free Hamiltonians of \(\mathcal{C}\) and  \(\mathcal{R}\) so that
\begin{equation}\label{eq:Ham1}
\hat{\mathcal{H}} = \hat{H}_\mathcal{R}\otimes \hat{1}_\mathcal{C} + \hat{1}_\mathcal{R} \otimes \hat{H}_\mathcal{C},
\end{equation}
where \(\hat{H}_\mathcal{R}\) is the free Hamiltonian of the rest of the universe, \(\hat{H}_\mathcal{C}\) is that of the clock, and \(\hat{1}_\mathcal{C}\) and \(\hat{1}_\mathcal{R}\)  are the unit observables of the clock and the rest of the universe, respectively. This explains dynamical evolution because, although the composite system is stationary, these subsystems will `evolve' relative to each other, as we will see shortly.

The third and final assumption is that the clock is an ideal one so that it has an observable \(\hat{t}\) that is conjugate to its Hamiltonian, i.e.
\begin{equation} \label{eq:cannonical}
[\hat{t},\hat{H}_\mathcal{C}]=i\hat{1}_\mathcal{C}.
\end{equation}
We shall call  \(\hat{t}\)  the clock's `time observable' since its value represents the time as indicated by the clock. For instance, an eigenstate \(\ket{t}\) of \(\hat{t}\), which has eigenvalue \(t\), will have the unequivocal physical meaning `according to the clock, it is time \(t\)' or as a shorthand `it is clock time \(t\)'. We use the term \textit{clock time} to distinguish it from time as indicated by the time parameter \(\lambda\), as the two are markedly different notions of time. For instance, \(\hat{t}\) is a q-number (i.e.\ an operator), whereas \(\lambda\) is a real-valued parameter; moreover, a measurement of \(\hat{t}\) does not yield the value of \(\lambda\).

The algebra expressed in Eq.~\ref{eq:cannonical} determines the spectra of eigenvalues of \(\hat{t}\) and \(\hat{H}_C\); both their spectra are equal to \(\mathbb{R}\). As usual, the eigenstates of \(\hat{t} \) are mutually orthogonal, meaning they are, in principle, distinguishable from one another. Since \(\hat{t}\) has a continuous range of distinguishable eigenstates, the clock's dial has infinite resolution – this is an idealisation, as the dial of a real clock has finite resolution. While the Page-Wootters construction has also been extended to non-ideal clocks \cite{hohn2021trinity,chataignier_relational_2024,hausmann_measurement_2025}, we will use ideal clocks with the understanding that they approximate the behaviour of ever more precise finite-dimensional real clocks.
 Due to Eq.~\ref{eq:cannonical}, which states that \(\hat{t}\) and \(\hat{H}_\mathcal{C}\) are conjugate observables, it also follows that 
\begin{equation}
\hat{H}_\mathcal{C}\ket{t} = i \frac{\text{d}}{\text{d}t}\ket{t}.
\end{equation}

The three assumptions described in Eqs.~\eqref{eq:stationary}, \eqref{eq:Ham1}, and \eqref{eq:cannonical} can be used to demonstrate that the rest of the universe will evolve according to the Schrödinger equation relative to clock time: taking the inner product of the universal state vector and an eigenstate \(\ket{t}\) of the clock, one finds that 
\begin{align} \label{eq:evolution}
& \bra{t}\hat{\mathcal{H}}\kket{\Psi} = 0, \\
\leftrightarrow & \bra{t}\hat{H}_\mathcal{R}\otimes \hat{1}_\mathcal{C}\kket{\Psi} = -\bra{t}\hat{1}_\mathcal{R} \otimes \hat{H}_\mathcal{C} \kket{\Psi}, \\
\leftrightarrow & \hat{H}_\mathcal{R} \brakket{t}{\Psi} = i \frac{\text{d}}{\text{d}t}\brakket{t}{ \Psi},
\end{align}
where to get from the first to the second line, we use the expression for \(
\hat{\mathcal{H}}\) shown in Eq.~\eqref{eq:Ham1}. The term \( \ket{\psi(t)}\stackrel{\text{def}}{=} \brakket{t}{\Psi} \) is called the \textit{relative state} \cite{everett1973theory}, as it represents the state of the rest of the universe, \(\mathcal{R}\), relative to the clock being in state \(\ket{t}\). Eq.~\eqref{eq:evolution} implies that the relative state evolves according to the Schrödinger equation with respect to clock time since
\begin{equation} \label{eq:Schrödinger}
\hat{H}_\mathcal{R} \ket{\psi(t)} = i \frac{\text{d}}{\text{d}t}\ket{\psi(t)}.
\end{equation}
Hence, the Page–Wootters construction explains evolution in the apparent absence of the time parameter, \(\lambda\), since despite the state of the composite system being stationary with respect to \(\lambda\), \(\mathcal{R}\) changes relative to the clock \(\mathcal{C}\).\footnote{This is also how the Page–Wootters construction provides a non-circular explanation of change. That is, it allows us to explain change in terms of things that do not change since the universal state vector is stationary, yet it allows us to explain the change of the relative states, which change relative to a clock.} This gives physical meaning to change and time in quantum theory, which is why Kuypers \cite{kuypers2022quantum} suggests the Page–Wootters construction is more aptly called `the quantum theory of time', a name that is also preferable in that it encompasses the improvements to the original construction of Page \& Wootters.
The Page-Wootters construction has also been recently linked to the theory of `quantum reference frames', where the Page-Wootters clock can be interpreted as a `temporal quantum reference frame' for the rest of the universe \cite{de_la_hamette_perspective-neutral_2021,ali_ahmad_quantum_2022,hohn2021trinity,hohn_equivalence_2021,hoehn_quantum_2023}.
 
\section{The problem} \label{sec:problem}
Page \& Wootters made the important assumption that the clock and the rest of the universe are non-interacting \cite{page1983evolution}, which is useful because it ensures that the rest of the universe evolves unitarily relative to the clock, as shown in Eq.~\eqref{eq:Schrödinger}. However, if this were a necessary assumption for the model, it would imply that the clock cannot be measured, as measuring it requires an interaction between the clock and a measurer. That is problematic because if the clock really cannot be measured, it would be as unphysical as the time parameter \(\lambda\), making the clock's presence pointless. 

Other authors have noticed this problem as well; for instance, Adlam \cite{adlam2022watching} summarises it as follows: \textit{`Indeed it is common for applications of the [Page– Wootters] formalism to represent the clock as non-interacting \textellipsis no provision [is] made for observers to actually obtain time readings from the clock.'} 

In this work, we provide such a provision: we show that clocks are measurable within the Page–Wootters construction because the model allows for measurers (in particular, \textit{timers}) to read the time from the clock. For this, the assumption that the clock is non-interacting has to be relaxed, but this does not imply a violation of unitarity. As we will demonstrate, the clock and timer will invariably evolve unitarily relative to the clock time.

Other authors have also noted that the clock and the rest of the universe can interact, albeit in different contexts – see, for instance, Refs.~\cite{giovannetti2015quantum,smith2019quantizing,castro-ruiz2020quantum,smith2020quantum,rijavec2023robustness}. In particular, Ref.~\cite{smith2019quantizing} demonstrates the existence of Page–Wootters states in settings where the clock and the rest of the universe interact. What these other models do not show, however, is that such interactions can result in a measurement of the clock. 

There is a second problem with the Page–Wootters construction, related to the one mentioned above: in the original formulation, clocks in the everyday sense of the word – e.g. the clocks used in laboratory experiments – would not qualify as clocks in the Page–Wootters sense, because the former can be measured. This makes it unclear how the theoretical description of clocks proposed by Page \& Wootters relates to actual clocks and to practical theories of time and timekeeping. Since we introduce a theory of clock measurement, our extension allows timekeeping devices (both natural and artificial ones) to satisfy the definition of Page–Wootters clocks, unifying the construction with practical theories of clocks and time, such as those used in experimental settings. This connection also renders the construction more amenable to empirical testing, by linking it to real-world clocks on which measurements can be performed.

\section{Measuring time} \label{sec:time}
We will show that the clock can be measured. Here, we rely on the theory of measurement as developed by Everett and others \cite{everett1973theory, kuypers2021everettian}, which allows us to treat a measurer of the clock as a quantum system and measurement as a unitary process. These are important assumptions for the Page-Wootters construction, as the construction requires that systems evolve unitarily.

In particular, we will demonstrate that the Page-Wootters construction allows for a \textit{timer} to measure the clock. We define a timer as a system that can distinguish between two specified time intervals—namely, the interval before and the interval after the measurement interaction. More sophisticated clock measurers can be built from collections of such timers. In our model, the measurement takes a finite amount of time, causing there to be a slight overlap between the intervals before and after the interaction. This overlap is unproblematic, as it can be made arbitrarily small.

The timer is modelled as a quantum system, and a measurement is defined as a certain type of interaction between the timer and the clock. Before the measurement interaction occurs, the timer is assumed to be at rest in some initial state; then, the interaction must be such that, after it ends, the timer is at rest again in a state orthogonal to the initial one. In this way, the two orthogonal states encode one bit of information about the time on the clock: if the timer is found in the initial state, the measurement has not yet occurred; and if it is found in the orthogonal state, the interaction has already taken place. When we speak of `before' and `after' the interaction, we mean with respect to clock time since the model as a whole is stationary with respect to the time parameter~\(\lambda\). Subsequent measurements of the timer can then be used to transfer this information about the clock to other systems.

In the original Page–Wootters construction, the clock is an ideally isolated system. Hence, it cannot be observed, as explained in Sec.~\ref{sec:problem}. To solve this problem, we present a variant of the construction in which the clock can be observed because we allow for an interaction between it and a timer. Our approach is based in part on Deutsch's work on the Page–Wootters construction  \cite{deutsch1990measurement}.\footnote{Deutsch's model consists of a stationary state in which one qubit measures another qubit. In Deutsch's model, this measurement of one qubit by another starts and ends at specific times, which is why the model has to allow for an interaction with the clock.} While in the original Page–Wootters construction, the clock and the rest of the universe are non-interacting, in both our and Deutsch’s models there is an interaction between these systems, and in both models, the interaction lasts only for a finite time. Specifically, we assume that the Hamiltonian of the model universe is
\begin{equation} \label{eq:Ham2}
\hat{\mathcal{H}}' = \hat{H}_\mathcal{R}\otimes \hat{P}_{\scriptscriptstyle\left[0,T\right]} + \hat{1}_\mathcal{R} \otimes \hat{H}_\mathcal{C},
\end{equation}
The product \(\hat{H}_\mathcal{R}\otimes \hat{P}_{\scriptscriptstyle\left[0,T\right]} \) represents the aforementioned interaction term, which is typically prohibited in the Page–Wootters construction. Here \(\hat{P}_{\scriptscriptstyle\left[0,T\right]}\) is a projector, which we define as follows:
\begin{align}
\hat{P}_{\scriptscriptstyle\left[0,T\right]} \stackrel{\text{def}}{=} \int_0^T \text{d}t\,\ket{t}\bra{t}.
\end{align}
So, the effect of this interaction will be that during the period \(0\leq t\leq T\), the rest of the universe will evolve according to the Hamiltonian \(\hat{H}_\mathcal{R}\), whereas the Hamiltonian of the rest of the universe is zero before and after this period. As will become apparent below, the presence of this interaction does not alter the main features of the model: the universal state vector will still be an eigenstate of \(\mathcal{\hat{H}}'\), and systems in that stationary universe will evolve unitarily relative to the clock.

Let us assume that the rest of the universe consists of a single qubit, \(\mathcal{Q}\), whose four basic observables are the Pauli matrices, \(\hat{\sigma}_x\), \(\hat{\sigma}_y\), \(\hat{\sigma}_z\), and the unit observable \(\hat{1}_\mathcal{R}\). These observables of the qubit satisfy the Pauli algebra, as their name suggests: that is to say that \([\hat{\sigma}_x, \hat{\sigma}_y ] = i 2 \hat{\sigma}_z \), as well as cyclic permutations of that expression over \(x\), \(y\), and \(z\), and \(\hat{\sigma}_x^2=\hat{\sigma}_y^2=\hat{\sigma}_z^2=\hat{1}_\mathcal{R}\). Of particular importance will be the eigenstates of \(\hat{\sigma}_z\), which are \(\ket{1}\) and \(\ket{0}\). \(\mathcal{Q}\)  will function as a simple timer, in that \(\mathcal{Q}\) will store exactly one bit of information about what time it is according to the clock because \(\mathcal{Q}\) will store an answer to the question, `is it earlier than \(t=T\) or later than \(t=0\)?' It does so since the qubit stores evidence of its interaction with the clock, marking that period. 

Before its interaction with the clock, we will assume \(\mathcal{Q}\) to be at rest in the \(\ket{0}\) state. We will let the interaction between the qubit and the clock be such that, following their interaction, which ends at \(t=T\), the qubit's eigenstate will have been toggled to \(\ket{1}\). Therefore, if one were to find the qubit in state \(\ket{1}\), this would be evidence that the qubit and the clock have interacted, implying it must be later than \(t=0\). To achieve this change in the state of \(\mathcal{Q}\), let us assume the Hamiltonian of the qubit is
\begin{equation}
\hat{H}_{\mathcal{R}} = \frac{ \pi (\hat{1}_{\mathcal{R}}-\hat{\sigma}_x)}{2T}.
\end{equation}
The effect of this Hamiltonian after \(T\) units of time is equivalent to that of a NOT gate since \(\exp(-i T \hat{H}_{\mathcal{R}})=\hat{\sigma}_x\) is the unitary that implements the NOT gate on \(\mathcal{Q}\). Consequently, the interaction term \(\hat{H}_\mathcal{R}\otimes \hat{P}_{\scriptscriptstyle\left[0,T\right]}\) will toggle the state of the qubit, as desired.

For the above-described history to be a permitted one in the Page–Wootters construction, there must be a universal state vector that is an eigenstate of the Hamiltonian of Eq.~\eqref{eq:Ham2} so that this energy eigenstate describes that history in which the timer measures clock time. That is to say, there must be an eigenstate of the Hamiltonian in which the state of \(\mathcal{Q}\) is toggled during its interaction with the clock. That energy eigenstate is the following one:
\begin{equation} \label{eq:Phialpha}
\kket{\Phi} \stackrel{\text{def}}{=} \int_{-\infty}^{+\infty}\text{d}t\,  \left( \theta(t<0) \ket{0}  + \theta(0\leq t \leq T) e^{ -i\hat{H}_{\mathcal{R}} t} \ket{0}  + \theta(t>T) \ket{1}  \right) \ket{t},
\end{equation}
where the function \(\theta\) equals \(1\) for the interval shown in \(\theta\)'s argument, whilst \(\theta\) is zero otherwise. Evidently, the qubit is at rest in state \(\ket{0}\) at clock times \(t<0\), but during the period \(0\leq t \leq T\), it interacts with the clock so that, at time \(t>T\), it is at rest again in the \(\ket{1}\) state.

To prove that \(\kket{\Phi}\) is an energy eigenstate of \(\hat{\mathcal{H}}'\), let us first consider how the interaction term acts on the state vector,
\begin{equation}
     \left( \hat{H}_{\mathcal{R}}\otimes \hat{P}_{\scriptscriptstyle\left[0,T\right]}\right) \kket{\Phi}  = \int_{0}^{T}\text{d}t\,  \hat{H}_{\mathcal{R}} e^{ -i\hat{H}_{\mathcal{R}} t} \ket{0}  \ket{t} ,
\end{equation}
which can alternatively be expressed as
\begin{multline}\label{eq:simplifiedaction}
\left( \hat{H}_{\mathcal{R}}\otimes \hat{P}_{\scriptscriptstyle\left[0,T\right]}\right) \kket{\Phi} =  i \int_{-\infty}^{+\infty}\text{d}t\, \frac{\text{d}}{\text{d}t} \left( \theta(t<0) \ket{0}  
\vphantom{e^{ -i\hat{H}_{\mathcal{Q}} t}} \right.\\
\left. + \theta(0\leq t \leq T) e^{ -i\hat{H}_{\mathcal{R}} t} \ket{0}  + \theta(t>T) \ket{1}  \right) \ket{t} .
\end{multline} 
Note that the function \(\theta\) is discontinuous, and its derivative with respect to \(t\) is, in general, a sum of Dirac delta functions; the delta-function terms that result from these derivatives of \(\theta\) in the expression above all cancel. Using Eq.~\eqref{eq:simplifiedaction}, one can readily verify that the effect of \(\hat{\mathcal{H}}'\) on \(\kket{\Phi}\) is
\begin{multline}
    \hat{\mathcal{H}}' \kket{\Phi} = i \int_{-\infty}^{+\infty} \text{d}t\, \frac{\text{d}}{\text{d}t} \left[ \left( \theta(t<0) \ket{0}   
    \vphantom{e^{ -i\hat{H}_{\mathcal{R}} t}} 
    \right. \right. \\ \left. \left. + \theta(0\leq t \leq T) e^{ -i\hat{H}_{\mathcal{R}} t} \ket{0}  + \theta(t>T) \ket{1}  \right ) \ket{t} \right ],
\end{multline}
which simplifies to
\begin{equation}\label{eq:eigenstate_condition}
      \hat{\mathcal{H}}' \kket{\Phi}  =   0.
\end{equation}
Here, we use integration by parts and the fact that the boundary terms at infinity vanish, as proven in Appendix \ref{app:RiggedHilbertspace}. Eq.~\eqref{eq:eigenstate_condition} concludes our proof that \(\kket{\Phi}\) is an energy eigenstate of  \(\mathcal{H}'\), specifically one with zero energy, making it a bona fide Page-Wootters state. 

Our main result is, thus, as follows: \(\kket{\Phi}\) is a Page–Wootters state in which a timer, namely \(\mathcal{Q}\), interacts with the clock, thereby measuring clock time, showing that the Page–Wootters construction allows the clock to be measured. Crucially, the timer evolves unitarily relative to the clock, so unitary evolution is evidently preserved despite the interaction between the two systems.

In our model, the ‘rest of the universe’ consists of a single system, namely the timer. Additional systems can be straightforwardly included by assuming they do not interact with either the clock or the timer. This adds a separate term to the Hamiltonian in Eq.~\eqref{eq:Ham2}, without introducing any interaction terms, and one would then find that the additional system evolves relative to clock time, as in the standard Page–Wootters construction, while the timer measures that time. Moreover, one can include systems that interact with the timer after it has measured the clock, allowing information about the clock to flow from the timer to those systems. This is because one system can measure another in the Page–Wootters construction, as shown by Deutsch \cite{deutsch1990measurement}, whose model partly underlies ours.

\subsection{Did the timer time the time?}\label{sec:did}
During this simple measurement of a clock, the qubit acts as a timer since its state contains information about that of the clock. Specifically, relative to the qubit being in the state \(\ket{0}\), the clock is never in an eigenstate \(\ket{t}\) with eigenvalue \(t \geq T\). This can be demonstrated by using the expectation values of the following projectors:
\begin{equation}\label{eq:projectors}
\Pi_{0}(\hat{\sigma}_z)\stackrel{\text{def}}{=}\ket{0}  \bra{0} \otimes \hat{1}_\mathcal{C}, \quad \Pi_{1}(\hat{\sigma}_z)\stackrel{\text{def}}{=}\ket{1}  \bra{1} \otimes \hat{1}_\mathcal{C}, \quad \Pi_t (\hat{t})\stackrel{\text{def}}{=} \hat{1}_\mathcal{R} \otimes \ket{t}  \bra{t}.
\end{equation}
Eq.~\ref{eq:projectors} above defines three projectors, each labelled with a subscript that is an eigenvalue of the q-number in the projector's argument; these subscripts correspond to the eigenstates that the projectors project onto. We use these definitions to simplify the expressions below. Given the state vector \(\kket{\Phi}\), as defined in Eq.~\eqref{eq:Phialpha}, one can readily verify that
\begin{align} 
  \bbra{\Phi }\Pi_{0}(\hat{\sigma}_z) \Pi_t (\hat{t})\kket{\Phi} = 0 \quad \text{for all} \quad t\geq T,  \label{eq:projectorsone}  \\
  \bbra{\Phi}\Pi_{0}(\hat{\sigma}_z) \Pi_t (\hat{t})\kket{\Phi} \not= 0 \quad \text{for all} \quad t < T.  \label{eq:projectorstwo}
\end{align}
This demonstrates our earlier claim that the clock is never in a state \(\ket{t}\) with \(t \geq T\) if the measurer is in state \(\ket{0}\). We include that the expectation value of the projector \(\Pi_{0}(\hat{\sigma}_z) \Pi_t (\hat{t})\) is non-zero prior to clock time \(T\) to demonstrate that, relative to the clock, the state of the qubit is different before and after the interval \(0\leq t \leq T\). As is evident from Eqs. \eqref{eq:projectorsone} and \eqref{eq:projectorstwo}, when the qubit is in the state \(\ket{0}\), it is certain that the time, as represented by the clock, is \(t<T\). 

Likewise, when the qubit is in the state \(\ket{1}\), it is certain that the time, as represented by the clock, is \(t>0\) because
\begin{align}
  \bbra{\Phi} \Pi_{1}(\hat{\sigma}_z) \Pi_t (\hat{t})\kket{\Phi} = 0 \quad \text{for all} \quad t<0,\\
  \bbra{\Phi} \Pi_{1}(\hat{\sigma}_z) \Pi_t (\hat{t}) \kket{\Phi} \not= 0 \quad \text{for all} \quad t \geq 0.
\end{align}
Notably, there is a slight ambiguity in the timer since, during the period \(0 < t < T\), the qubit may be found in either state \(\ket{0}\) or \(\ket{1}\). This must be a general feature of any real timer: a timer will always need to change its state continuously during a finite period. For the qubit to be a good timer, the period \(T\) must be small relative to the relevant time scales so that \(T\) is negligible. In principle, \(T\) can always be made small enough to be irrelevant, or put differently,  there is no physical limit, short of perfection, on how closely a timer can approach the ideal case in which \(T=0\) because there is no bound on how small \(T\) can be made. This is so because there is no bound on how fast a gate, such as the NOT gate, can be enacted, as shown in \cite{deutsch1982there}. In Appendix \ref{app:inst}, we discuss an idealisation of this measurement procedure that is truly instantaneous. Unlike the more realistic model described here, this idealisation requires a non-Hermitian interaction term.

The qubit contains precisely \(1\) bit of information about the time on the clock since the qubit contains an answer to the binary question, `Does the clock indicate it is earlier than \(t=T\) or later than \( t= 0 \)?' By including more qubits, the timer can be made more sophisticated so that it can, for example, record whether it is earlier or later than the times within a second interval. By including an arbitrary number of qubits, the timer can measure clock time arbitrarily well. In fact, as we will show in the next section, the timer can be idealised.

\subsection{Synchronising two clocks}
The timer described in Sec.~\ref{sec:time} is a simple device, consisting of a single qubit. It can be made more sophisticated by including additional qubits, which allow it to more accurately measure clock time. Here we show that there is no bound on how sophisticated the timer can be made because an ideal clock can be used to measure the time on another ideal clock, which provides a good model for a timer consisting of an arbitrarily large number of qubits.

To that end, let us consider a universe consisting of two ideal clocks, \(\mathcal{C}_1\) and \(\mathcal{C}_2\), which have the respective pairs of canonically conjugate observables \((\hat{t}_1\otimes \hat{1}_{\mathcal{C}_2},\hat{H}_{\mathcal{C}_1}\otimes \hat{1}_{\mathcal{C}_2})\) and \((\hat{1}_{\mathcal{C}_1} \otimes \hat{t}_2 ,\hat{1}_{\mathcal{C}_1} \otimes \hat{H}_{\mathcal{C}_2} )\).  Note that the observables of  \(\mathcal{C}_1\) commute with those of \(\mathcal{C}_2\) because the Hilbert space of the composite system is a tensor product of their individual Hilbert spaces, as usual. Our current question is whether an ideal clock, \(\mathcal{C}_1\), can be used to read the time on another ideal clock, \(\mathcal{C}_2\). The answer is: yes.

For this analysis, we assume that the Hamiltonian of the model universe is
\begin{align}
\hat{\mathcal{H}}''_{t_0}=  \hat{H}_{\mathcal{C}_1} \otimes \hat{P}_{\scriptscriptstyle\left[t_0,\infty\right)} + \hat{1}_{\mathcal{C}_1} \otimes \hat{H}_{\mathcal{C}_2}
\label{eq:H_sync}
\end{align}
with
\begin{align}
\hat{P}_{\scriptscriptstyle\left[t_0,\infty\right)}  \stackrel{\text{def}}{=}  \int_{t_0}^{\infty}\text{d}t\, \ket{t}\bra{t}.
\end{align}
The effect of this interaction term will be to synchronise the two clocks with one another so that, from \(t_0\) onwards, the clocks will be ticking in lockstep.\footnote{Usually, we consider two clocks synchronised if they intrinsically tick with the same speed (in the same frame of reference) and if their counters have been set to the same value at some point. The two clocks will then remain synchronised after this event as long as they are in the same reference frame. Here, the first clock is perpetually recording the time on the second one.}
The following energy eigenstate of \(\hat{\mathcal{H}}_{t_0}''\)  reflects a history in which the two ideal clocks become synchronised:
\begin{equation}
    \kket{\Theta_{t_0}} \stackrel{\text{def}}{=}  \int_{-\infty}^{+\infty}\text{d}t\, \left( \theta(t < t_0) \ket{0}  + \theta(t\geq t_0) \ket{t}  \right) \ket{t} ,
\end{equation}
where \(\ket{0}\) is the eigenstate of \(\hat{t}_1\) with eigenvalue zero. It can be readily verified, in the same manner as that of the previous sections, that \(\kket{\Theta_{t_0}}\) is an eigenstate of \(\hat{\mathcal{H}}_{t_0}''\) since
\begin{align}
    \hat{\mathcal{H}}''_{t_0} \kket{\Theta_{t_0}} & = i  \int_{-\infty}^{+\infty} \text{d}t\,  \frac{\text{d}}{\text{d}t} \left[ \left( \theta(t < t_0) \ket{0}  + \theta(t \geq t_0) \ket{t}  \right) \ket{t} \right]   = 0,
\end{align}
where we again use the result that the boundary terms vanish, as proven in Appendix \ref{app:RiggedHilbertspace}.

The state \(\kket{\Theta_{t_0}}\) describes a system consisting of two clocks. The first of these clocks, \(\mathcal{C}_1\), is initially at rest in the state \(\ket{0}\) relative to \(\mathcal{C}_2\). But when this second clock, \(\mathcal{C}_2\), is in state \(\ket{t_0}\), there is an interaction between both clocks that prompts \(\mathcal{C}_1\) to start ticking in lockstep with \(\mathcal{C}_2\). This is an idealisation of the previously analysed timer since no time was needed to synchronise \(\mathcal{C}_1\) and \(\mathcal{C}_2\), whereas the timer needed a finite period to perform its measurement. Moreover,  \(\mathcal{C}_1\) has infinitely many distinguishable states since it has infinitely many orthogonal eigenstates \(\{\ket{t} : \hat{t}_1 \ket{t} = t\ket{t}, t\in \mathbb{R}\}\), so \(\mathcal{C}_1\) can record time arbitrarily well, whereas the qubit was only able to store one bit of information about the time.

\(t_0\) marks the time the two clocks become synchronised. Consider letting \(t_0\rightarrow -\infty\) so that the synchronisation procedure occurs in the infinite past. In this case, the Hamiltonian of the universe reduces to one without interactions since
\begin{equation} \label{eq:int_perf_sync}
    \lim_{t_0 \rightarrow -\infty} \hat{\mathcal{H}}''_{t_0} = \hat{H}_{\mathcal{C}_1} \otimes \hat{1}_{\mathcal{C}_2}  + \hat{1}_{\mathcal{C}_1} \otimes \hat{H}_{\mathcal{C}_2},
\end{equation}
and in that same limit, the zero-energy eigenstate becomes
\begin{equation} \label{eq:perf_sync}
    \lim_{t_0 \rightarrow -\infty} \kket{\Theta_{t_0}} = \int_{-\infty}^{+\infty}\text{d}t\, \ket{t} \ket{t}.
\end{equation}
Eq.~\eqref{eq:int_perf_sync} no longer contains an interaction term between \(\mathcal{C}_1\) and \(\mathcal{C}_2\), so the Hamiltonian of the universe describes two non-interacting clocks. Nevertheless, the two clocks are perfectly synchronised, as can be seen in Eq.~\eqref{eq:perf_sync}. This is because the clocks became synchronised in the infinite past, so no interaction is required at any finite time to synchronise them. Put differently, in this limit it is a feature of the state of the universe that the clocks have always been and always will be ticking in lockstep with one another. This is why states like those in Eq.~\eqref{eq:perf_sync} are for example used in Ref.~\cite{maccone2020quantum} to demonstrate that Page–Wootters allows for readings of the clock in the absence of interactions with the clock since these models assume the existence of two good clocks that were synchronised in the infinite past. Such models are, therefore, encompassed in our more general framework of clock measurements and synchronisation.

\section{Conclusions and new problems}
We have presented a model of a stationary universe that features a dynamical measurement of a clock following the Page–Wootters construction. Our model consists of a clock and a timer, and unlike the original construction, we allow for an interaction between the two, which enables the timer to read the time from the clock. This demonstrates that the time of the theoretical clock in the Page-Wootters construction can be accessed. Hence, the clock is physical since it can be measured, just like a real clock. 

Crucially, the type of timer-clock interaction we have considered leads to a unitary dynamics, thus preserving the standard dynamical evolution of quantum theory. This was not guaranteed since other types of interactions can lead to a non-unitary dynamics \cite{paiva_flow_2022,paiva_non-inertial_2022}, and only specific classes of interactions lead preserve unitarity \cite{rijavec_conditions_2025}.

Our results are derived for ideal clocks, which are systems with infinite-dimensional Hilbert spaces. Such clocks serve as useful approximations to realistic ones, which have finite-dimensional Hilbert spaces \cite{hohn2021trinity}. Realistic clocks have been studied extensively (see, for instance, Refs.~\cite{chataignier_relational_2024, hausmann_measurement_2025, woods_autonomous_2019}). We expect our results to extend straightforwardly to this setting. We shall leave a detailed investigation for future work.

We also discussed a simple model of clock synchronisation, where one ideal clock synchronises with another through an interaction. After this interaction, the clocks remain synchronised and tick in unison indefinitely, and in a limiting case of this model, the clocks are naturally synchronised without any interaction. 

In the real world, the structure of spacetime causes two sychronised clocks to lose synchronisation if one of them starts moving relative to the other. The Page–Wootters construction does not account for this, as it does not describe space, so the clocks cannot move. By extending the model to include space, one could investigate the effects that relative motion would have on Page–Wootters clocks and their synchronisation, providing insights into the structure of spacetime and its relation to quantum theory. The results of this paper should serve as a stepping stone for such research, which we will leave for future study.

\section*{Acknowledgements}
Both authors are grateful to David Deutsch, Eric Marcus, and Max Velthoven for their feedback on earlier drafts of this paper. Their insights have greatly enhanced the quality of this work.

Sam Kuypers wishes to thank Stefan Wolf for invaluable discussions and for providing the opportunity to work on this project. He was supported in part by the Swiss National Science Foundation (SNSF) under grant number 200020\_182452 and by the Québec–Ontario Consortium on Quantum Protocols (QUORUM). He also wishes to thank Maria Santos for her moral support throughout the project.


\bibliographystyle{unsrtnat}
\bibliography{mybib}

\appendix 

\section{The Rigged Hilbert Space} \label{app:RiggedHilbertspace}
There is a technical issue in the Page–Wootters construction that is relevant to our work. The issue is that the universal state vector of Sec.~\ref{PWconstruction}, namely \(\kket{\Psi}\), is not an element of the Hilbert space of the universe, \(\mathscr{H}_U\). This is because we are considering an ideal clock and so the zero-energy eigenvalue of \(\hat{\mathcal{H}}\) lies in the continuum part of the spectrum, making \(\kket{\Psi}\) non-normalisable according to the standard inner product on \(\mathscr{H}_U\), usually called the \textit{kinematical} inner product \cite{rovelli_quantum_2004}. This can be demonstrated as follows. By using the resolution of the identity on the Hilbert space of the clock $\mathscr{H}_\mathcal{C}$ in terms of the eigenstates of \(\hat{t}\), we can write the universal state vector as
\begin{equation}\label{eq:decomposition}
    \kket{\Psi} = \int_{-\infty}^{+\infty}\text{d}t\, \ket{\psi(t)} \ket{t}.
\end{equation}
Let us assume that the relative state \(\ket{\psi(t)}\) is normalised at time \(t=0\); the relative state obeys the norm-preserving Schrödinger equation  \eqref{eq:Schrödinger}, so it is automatically normalised at all other times, too, viz. \(\braket{\psi(t)}{\psi(t)} = 1\) for all \(t\in\mathbb{R}\). Consequently, \(\kket{\Psi}\) does not have a finite norm and is, therefore, not part of the Hilbert space \(\mathscr{H}_U\). 

This problem is solved by considering a different type of inner product on  \(\mathscr{H}_U\), namely, the \textit{physical} inner product defined as \cite{rovelli_quantum_2004,smith2019quantizing}:
\begin{equation}
    \bbrakket{\Phi}{\Psi}_{phy}=\bbraket{\Phi}{t}\brakket{t}{\Psi},
\end{equation}
for any choice of \(t\). The physical inner product does not depend on \(t\) unless the clock interacts with the rest of the universe in a way that leads to a non-unitary evolution.
According to the physical inner product, \(\kket{\Psi}\) is normalisable, solving the issue described above.

Nevertheless, there are further subtleties arising from the use of ideal clocks, similar to the problems that one encounters with the (generalised) eigenstates of position or momentum for a free particle. The rigorous way to treat these states mathematically employs the so-called \textit{rigged Hilbert space} formalism  \cite{gelfand1964}. This construction is what justifies, in the first place, the resolution of the identity in terms of the eigenstates of \(\hat{t}\) and the notion of integral over state vectors in Eq.~\eqref{eq:decomposition}. We briefly and non-rigorously review this formalism here, but we refer the reader to Refs.~\cite{gelfand1964,de_la_Madrid_2005} for extensive discussions.

A subtlety that is central to this work arises due to the fact that \(\kket{\Psi}\) is part of the rigged Hilbert space; the subtlety in question is that, using Eq.~\eqref{eq:decomposition}, we have
\begin{equation}
    \hat{\mathcal{H}}\kket{\Psi} = i \int_{-\infty}^{+\infty}\text{d}t\, \frac{\text{d}}{\text{d}t} \left(\ket{\psi(t)} \ket{t} \right) = i\left.\vphantom{\int} \ket{\psi(t)} \ket{t} \right|_{-\infty}^{+\infty},
    \label{eq:boundary_terms}
\end{equation}
where the last step can be seen as the result of an integration-by-parts rule. The norm of the relative state \(\ket{\psi(t)}\) is preserved so that these boundary terms are apparently non-zero. Yet, from our earlier analysis, we know that \(\kket{\Psi}\) is an energy eigenstate with zero energy, implying that the boundary terms should vanish. How can this be?

The solution to this problem is that in the rigged Hilbert space formalism, \(\kket{\Psi}\) has a formal definition as a functional acting on so-called test vectors. These test vectors are well-behaved in that they are infinitely differentiable and go to zero rapidly at infinity. Because they go to zero at infinity, the inner product of such test vectors with the boundary terms in Eq.~\eqref{eq:boundary_terms} will invariably vanish, so the boundary terms are formally equivalent to zero in the rigged Hilbert space. Hence, \(\kket{\Psi}\) is an exact energy eigenstate with zero energy, solving the technical problem.

To define the aforementioned test vectors, let us consider the Hilbert space $\mathscr{H}_{\mathcal{C}}=L^2(\mathds{R})$ and the algebra $\mathfrak{h}$ generated by $\hat{H}_{\mathcal{C}}$ and $\hat{t}$ such that $[\hat{t},\hat{H}_{\mathcal{C}}]=i \hat{1}_{\mathcal{C}}$ (in this case these operators are defined by their action on the functions in $L^2(\mathds{R})$).
We can consider the space of test vectors $\ket{\chi_{\mathcal{C}}}$ such that $\norm{\hat{O}_{\mathcal{C}}\ket{\chi_{\mathcal{C}}}}<\infty$ for any $\hat{O}_{\mathcal{C}} \in \mathfrak{h}$. This will be the space of functions in $L^2(\mathds{R})$ with partial derivatives rapidly decreasing at infinity (Schwartz space). If we call $\Phi$ the space of all such test vectors, we can consider its dual $\Phi^*$, that is, the space of all the linear functionals on $\Phi$.  $\Phi^*$ is usually called the space of `distributions' and we have that $\Phi\subset\mathscr{H}_{\mathcal{C}}\subset\Phi^*$ \cite{gelfand1964}.
We can denote the members of $\Phi^*$ with the bra notation as $\bra{\omega_{\mathcal{C}}}$, so that $\braket{\omega_{\mathcal{C}}}{\chi_{\mathcal{C}}}\in\mathds{C}$ for any test vector $\ket{\chi_{\mathcal{C}}}$.

$\Phi^*$ contains the so-called `generalised eigenvectors' of $\hat{H}_{\mathcal{C}}$ and $\hat{t}$, defined as the bras $\bra{\varepsilon}$ and $\bra{t}$ such that $\bra{\varepsilon}\hat{H}_{\mathcal{C}}\ket{\chi_{\mathcal{C}}}=\varepsilon \braket{\varepsilon}{\chi_{\mathcal{C}}}$ and $\bra{t}\hat{t}\ket{\chi_{\mathcal{C}}}=t \braket{t}{\chi_{\mathcal{C}}}$ for every $\ket{\chi_{\mathcal{C}}}\in \Phi$, respectively. By taking their conjugate, we get the eigenstates $\ket{\varepsilon}$ ($\ket{t}$) of $\hat{H}_{\mathcal{C}}$ ($\hat{t}$) that we have used in this paper. Within the Rigged Hilbert space formalism, we can also give precise meaning to expressions such as $\int \text{d}t \, \ket{\varepsilon}\bra{\varepsilon} =\hat{1}_{\mathcal{C}}$ and $\ket{\chi_{\mathcal{C}}}=\int \text{d}t \, \braket{\varepsilon} {\chi_{\mathcal{C}}}\, \ket{\varepsilon}$ \cite{de_la_Madrid_2005}.

Crucially, the generalised eigenvectors must be understood as functionals on $\Phi$. Therefore, expressions such as the one in Eq.~\eqref{eq:boundary_terms}
\begin{equation}
    \hat{\mathcal{H}}\kket{\Psi} = i\left.\vphantom{\int} \ket{\psi(t)} \ket{t} \right|_{-\infty}^{+\infty},
    \label{eq:boundary_terms2}
\end{equation}
make sense only when they act on test vectors.
Assuming that \(\mathscr{H}_\mathcal{R}\) is finite-dimensional, the test vectors have the following general expression:
\begin{equation} \label{eq:test}
\kket{\chi} \stackrel{\text{def}}{=} \int_{-\infty}^{+\infty} \text{d}t\,\sum_j\mu_j(t) \ket{\phi_j} \ket{t},
\end{equation}
 where \(\left\{\ket{\phi_j}\right\}_j\) is an orthonormal basis of \(\mathscr{H}_\mathcal{R}\), and each \(\mu_j(t)\) is an infinitely differentiable, square-integrable function. The inner product of any such test vector with \(\hat{\mathcal{H}}\kket{\Psi}\) will vanish because the functions \(\mu_j(t)\) tend to zero at infinity:
\begin{equation}
    \bbra{\chi}\hat{\mathcal{H}}\kket{\Psi} =  i  \left.\left(\textstyle \sum_j  \mu_j^*(t)\braket{\phi_j}{\psi(t)}   \right)\right|_{-\infty}^{+\infty}= 0,
\end{equation}
where we used Eq.~\eqref{eq:boundary_terms2} and the fact that any square-integrable function will tend to zero at infinity, i.e., \(\mu_j^*(t)\rightarrow 0\) when \(t \rightarrow \pm \infty\). Since the inner product of \(\hat{\mathcal{H}}\kket{\Psi}\) with \textit{any} test vector vanishes, and since  state vectors in the rigged Hilbert space are defined by their action on such test vectors, these boundary terms are equivalent to zero, vindicating the fact that \(\kket{\Psi} \) is an eigenstate with zero energy.

\section{Instantaneous measurements}\label{app:inst}
In Sec.~\ref{sec:did}, we said that there are no limits on how fast the clock can be measured. Consequently, the measurement can indefinitely \textit{approach} an ideal instantaneous measurement.
Here, we show that there is also a mathematical way to implement a truly instantaneous measurement in the Page-Wootters construction.

Let us consider as before a qubit \(\mathcal{Q}\) and a clock \(\mathcal{C}\) but now with a global (non-Hermitian) Hamiltonian
\begin{equation}
\hat{\mathcal{H}}''' = i\,\left(\hat{1}_{\mathcal{Q}}-\hat{G}^{\dagger}\right)\otimes \ket{t^*}\bra{t^*} + \hat{1}_{\mathcal{Q}}\otimes\hat{H}_{\mathcal{C}},
\label{eq:H_inst_meas}
\end{equation}
where \(\hat{1}_{\mathcal{Q}}\) is the identity on \(\mathcal{Q}\),  \(\hat{G}\) is a gate (unitary) on the qubit, and \(t^*\) is the time at which the gate is to be applied. 
The eigenstate of \(\hat{\mathcal{H}}''\) that describes the application of the gate \(\hat{G}\) at time \(t^*\) is
\begin{equation}
\kket{\Xi} = \int_{-\infty}^{+\infty} \text{d}t\,  \left ( \theta(t<t^*) \ket{0} + \theta(t\geq t^*) \hat{G}\ket{0}  \right ) \ket{t} .
\label{eq:state_inst_meas}
\end{equation}
If \(\hat{G}\) is, for example, a NOT gate, the qubit has effectively measured the state $\ket{t^*}$ of the clock instantaneously.

Let us check that \(\kket{\Xi}\) is actually an eigenstate of \(\hat{\mathcal{H}}''\). The interaction term acts on \(\kket{\Xi}\) in the following way:
\begin{align}
&i\left[\left(\hat{1}_{\mathcal{Q}}-\hat{G}^{\dagger}\right)\otimes \ket{t^*}\bra{t^*}\right] \kket{\Xi} = i  \left(\hat{G}-\mathds{1}_{\mathcal{Q}}\right)\ket{0}  \ket{t^*}\nonumber\\
&=i\int_{-\infty}^{+\infty} \text{d}t\,  \frac{\text{d}}{\text{d}t}\left ( \theta(t<t^*) \ket{0} + \theta(t\geq t^*) \hat{G}\ket{0}  \right ) \ket{t} ,
\end{align}
and, therefore,
\begin{equation}
    \hat{\mathcal{H}}''' \kket{\Xi}  = i  \int_{-\infty}^{+\infty} \text{d}t \frac{\text{d}}{\text{d}t} \left[ \left(  \theta(t < t^*) \ket{0}  + \theta(t \geq t^*)\hat{G}\ket{0}  \right) \ket{t} \right]   = 0,
\end{equation}
where the last equality follows from our discussion in Appendix \ref{app:RiggedHilbertspace}

We have thus implemented a truly instantaneous measurement in the Page-Wootters construction. The price we had to pay is having introduced a \textit{non-Hermitian} interaction term in the Hamiltonian of the universe:
\begin{equation}
     i\,\left(\hat{1}_{\mathcal{Q}}-\hat{G}^{\dagger}\right)\neq \left[i\,\left(\hat{1}_{\mathcal{Q}}-\hat{G}^{\dagger}\right)\right]^{\dagger},
\end{equation}
unless \((\hat{G}+\hat{G}^{\dagger})/2=\hat{1}_{\mathcal{Q}}\) which implies that \((\hat{G}-\hat{1}_{\mathcal{Q}})^2=0\) and thus \(\hat{G}=\hat{1}_{\mathcal{Q}}\) (i.e., no gate is applied).
The non-Hermiticity is due to the fact that an instantaneous change cannot be described as a continuous unitary evolution with a Hermitian Hamiltonian.

Is the non-Hermiticity of the interaction term problematic? Usually, the answer would be 'yes' since non-Hermitian operators can have non-real expectation values. In the Page-Wootters construction, however, the only allowed states of the universe are the zero-eigenstates \(\kket{\Psi_j}\) of the total Hamiltonian of the universe. For a general Hamiltonian \(\hat{\mathcal{H}_G}\stackrel{\text{def}}{=} \hat{H}_{\mathcal{R}}+\hat{H}_{\mathcal{C}}+\hat{V}\) with an interaction term \(\hat{V}\), all the possible expectation values of \(\hat{V}\) at the level of the universe are given by
\begin{equation}
    \bbra{\Psi_j}\hat{V}\kket{\Psi_j}=-\bbra{\Psi_j}\hat{H}_{\mathcal{R}}+\hat{H}_{\mathcal{C}}\kket{\Psi_j} \quad \forall j ,
\end{equation}
which are real since both \(\hat{H}_{\mathcal{R}}\)  and \(\hat{H}_{\mathcal{C}}\) are self-adjoint. Therefore, even if \(\hat{V}\) is non-Hermitian, all its possible expectation values at the level of the universe are real. If instead one considers the average expectation value of \(\hat{V}\) in a time interval \(\Delta t\) around \(t\), defined as 
\begin{equation}
   \overline{\eexpval{\scalebox{0.9}{$\hat{V}$}}}_t  = \frac{\int_{t-\Delta t/2}^{t+\Delta t/2}\text{d}t'\,\bbra{\Psi_j}\Pi_{t'}(\hat{t})\hat{V}\Pi_{t'}(\hat{t})\kket{\Psi_j}}{\int_{t-\Delta t/2}^{t+\Delta t/2}\text{d}t'\,\bbra{\Psi_j}\Pi_{t'}(\hat{t})\Pi_{t'}(\hat{t})\kket{\Psi_j}},
\end{equation} 
one can get some non-real values.
For example, using the interaction term of Eq.~\eqref{eq:H_inst_meas} and the state of Eq.~\eqref{eq:state_inst_meas} one gets
\begin{equation}
\overline{\eexpval{\scalebox{0.9}{$\hat{V}$}}}_t=\begin{cases} 
    i\frac{\left(1-\bra{0}\hat{G}^{\dagger}\ket{0}\right)}{\Delta t} & \text{if } t^* \in \left[t-\Delta t/2,t+\Delta t/2 \right], \\
    0 & \text{otherwise},
\end{cases} 
\end{equation}
which can be, in general, non-real. However, it is not clear whether these expectation values can be measured within the PW construction. We leave this problem open for future discussion.

Finally, if instead of the qubit \(\mathcal{Q}\) we consider another ideal clock \(\mathcal{C}_2\) with a pair of canonically conjugate observables \((\hat{t}_2,\hat{H}_{\mathcal{C}_2})\) and starting in the state \(\ket{t'}\), any change under the unitary \(\hat{G}=e^{-i\hat{H}_{\mathcal{C}_2}\delta t}\) will result in a distinguishable state \(\ket{t'+\delta t}=e^{-i\hat{H}_{\mathcal{C}_2}\delta t}\ket{t'}\). If \(\delta t\ll 1\), the interaction term is well approximated by
\begin{equation}
    i\,\left(\hat{1}_{\mathcal{C}_2}-\hat{G}^{\dagger}\right)=\hat{H}_{\mathcal{C}_2}\delta t + O(\delta t^2),
\end{equation}
which is, at first order in \(\delta t\), self-adjoint. An ideal clock can thus measure the time of another ideal clock instantaneously using Hamiltonians that are self-adjoint within this approximation.

\end{document}